\documentclass{emulateapj}
\usepackage{multirow}

\slugcomment{Not to appear in Nonlearned J., 45.}

\shorttitle{CO in dusty quasars}
\shortauthors{Krips et al.}

\begin{document}


\title{CO emission in optically obscured (type-2) quasars at redshifts
${\rm\lowercase{z}}\approx$~0.1-0.4\thanks{}}


\author{M. Krips, R. Neri and P. Cox }
\affil{Institut de Radio Astronomie Millim\'etrique (IRAM),
300, rue de la Piscine, Domaine Universitaire, 38406 Saint Martin
d'H\`eres, France} 
\email{krips@iram.fr,neri@iram.fr,cox@iram.fr}

\begin{abstract}
We present a search for CO emission in a sample of ten type-2 quasar
host galaxies with redshifts of z$\approx$0.1-0.4. We detect
CO(J=1--0) line emission with $\geq$5$\sigma$ in the velocity
integrated intensity maps of five sources.  A sixth source shows a
tentative detection at the $\sim$4.5$\sigma$ level of its CO(J=1--0)
line emission. The CO emission of all six sources is spatially
coincident with the position at optical, infrared or radio
wavelengths. The spectroscopic redshifts derived from the CO(J=1--0)
line are very close to the photometric ones for all five detections
except for the tentative detection for which we find a much larger
discrepancy. We derive gas masses of
$\sim$(2-16)$\times$10$^{9}$M$_\odot$ for the CO emission in the six
detected sources, while we constrain the gas masses to upper limits of
M$_{\rm gas}$$\leq$8$\times$10$^{9}$M$_\odot$ for the four
non-detections. These values are of the order or slightly lower than
those derived for type-1 quasars. The line profiles of the CO(J=1--0)
emission are rather narrow ($\lesssim$300~km~s$^{-1}$) and single
peaked, unveiling no typical signatures for current or recent merger
activity, and are comparable to that of type-1 quasars. However, at
least one of the observed sources shows a tidal-tail like emission in
the optical that is indicative for an on-going or past merging event.

We also address the problem of detecting spurious $\sim$5$\sigma$
emission peaks within the field of view.

\end{abstract}



\begin{deluxetable*}{c@{\,}lcccccc}
\centering
\tablewidth{0pt}
\tablecaption{Source list}
\tablehead{
\multicolumn{2}{c}{Source$^a$} & \colhead{Redshift$^{b}$} & \colhead{D$_L$$^c$} & \colhead{Scale} 
&  \colhead{$\alpha_{\rm J2000}$$^d$} & \colhead{$\delta_{\rm J2000}$$^d$}  
& \colhead{Type$^{a}$} \\
 &        &  $z_{\rm opt}$       & (Mpc) & (kpc/arcsec) &  (hh:mm:ss.s)   & (dd:mm:ss.s) }
\startdata 
SWIRE2 & J021638.21$-$042250.8  &  0.304 & 1565 & 4.5 & 02:16:38.20 &  $-$04:22:50.8 &  2   \\
SWIRE2 & J021909.60$-$052512.9  &  0.099 &  450 & 1.7 & 02:19:09.60 &  $-$05:25:12.4 &  2   \\  
SWIRE2 & J021939.08$-$051133.8  &  0.150 &  706 & 2.5 & 02:19:39.09 &  $-$05:11:33.8 &  2   \\  
SWIRE2 & J022306.74$-$050529.1  &  0.330 & 1722 & 4.7 & 02:23:06.70 &  $-$05:05:29.1 &  2   \\
SWIRE2 & J022508.33$-$053917.7  &  0.293 & 1499 & 4.3 & 02:25:08.30 &  $-$05:39:17.7 &  2   \\
SDSS & J092014.11+453157.3      & 0.403  & 2179 & 5.4 & 09:20:14.10 &  $+$45:31:57.0 &  2   \\
SDSS & J103951.49+643004.2      & 0.402  & 2173 & 5.4 & 10:39:51.50 &  $+$64:30:04.0 &  2   \\
SSTXFLS & J171325.1+590531      &  0.126 &  583 & 2.2 & 17:13:25.10 &  $+$59:05:31.0 &  2   \\  
SSTXFLS & J171335.1+584756      &  0.133 &  619 & 2.3 & 17:13:35.10 &  $+$58:47:56.0 &  1R  \\  
SSTXFLS & J172123.1+601214      &  0.325 & 1691 & 4.8 & 17:21:23.10 &  $+$60:12:14.0 &  2   \\   [-0.15cm]
\enddata

\tablenotetext{$a$}{taken from \cite{lacy07b} and \cite{zaka08}.}

\tablenotetext{$b$}{Optical redshifts (spectroscopically determined).}

\tablenotetext{$c$}{Assuming the following cosmology: $\Omega_{\rm
M}$=0.27, $\Omega_\Lambda$=0.73 and H$_0$=71~km~s$^{-1}$~Mpc$^{-1}$.}

\tablenotetext{$d$}{Positions taken from \cite{simp06} (radio),
\cite{lacy07b} (optical/IR) or \cite{zaka08} (optical/IR).}

\tablenotetext{$e$}{Type of AGN: 2= optically obscured AGN;
1R=reddened type-1 AGN (type-1= un-obscured AGN). }

\label{tab1}
\end{deluxetable*}

\section{Introduction}

Molecular gas is an important diagnostic tracer to study the activity
processes in galaxies, whether in form of an accreting super-massive
central black hole (AGN) or intense starbursts (SB) or both. It does
not only build up the fuel for the activity but also traces essential
physical properties of the host galaxy, such as the kinematics,
distribution, excitation conditions and chemistry of its gas. As
activity occurs in many different types of galaxies and at various
levels of intensity and galaxy evolution, a big effort is put into
elucidating the physics behind its different occurrences.

Given the elementary importance of molecular gas, numerous studies
have been conducted to detect and map its distribution and kinematics
in galaxies. Most of them were focussed on sources in the local
universe given their easier detectability. However, recent technical
improvements make it now possible to probe the higher redshift galaxy
population
\cite[e.g.,][]{wanga11,riech11,riech09,tacc10,tacc08,tacc06,dadd10,
dadd09,grev05,neri03}. This is especially important because most of
the active galaxies, such as the ultra-luminous infrared
(ULIRGs)/sub-millimeter galaxies and quasars, are found in the
high-redshift universe emphasizing the tight correlation of the
activity of a galaxy and its evolution. However, it is still poorly
understood whether the nature of the activity plays a role for the
evolution of a galaxy and how; most current theories assume that both
activity types mark a special stage of evolution in a galaxy's life
\citep[e.g.,][]{san88a,san88b,hopk06a}. It is thus surprising that
molecular gas has been extensively studied in a large number of
(SB-powered) ULIRGs/submillimeter galaxies, while the (AGN-powered)
quasar population still lacks a similarly high number of systematic
studies. This lack of information is hence a shortcoming in current
theories of galaxy evolution.

While most recent molecular gas studies of quasars at redshifts of
z$<$0.5 concentrate on optically selected sources \cite[type-1
quasars; e.g.,][]{wink97,evan01,scov03,krip05a,krip07,bert07},
evidence has recently been presented that optically obscured
(type-2/reddened type-1) quasars may be found at least in equal
amounts \cite[at least for $z\lesssim$4:] []{reys08,lacy07a,mart06}.
Identifying type-2 quasars has been proven to be extremely difficult
because of their high obscuration at most wavelengths but success
rates have been significantly improved by several groups in the past
few years using the less obscured radio, X-ray, and/or Mid-Infrared
bands \cite[e.g.,][]{lacy07b,zaka08}.

The study of both, type-1 and type-2 quasars, is critical as it
challenges the viewing angle and merger-driven unification theories
for quasars.  While in the viewing-angle unification an edge-on clumpy
torus blocks the direct view onto type-2 quasars (Antonucci 1993), the
merger-driven unification assumes that type-2 quasars are in an
earlier stage of a merging event than type-1 quasars
\cite[e.g.,][]{hopk06a}. The latter scenario finds support by recent
observational results: 1.)  type-2 quasars are found to have higher
median star formation luminosities than type-1 quasars suggesting that
type-2 quasar activity goes along with an intensified star formation
activity in the host galaxy \cite[e.g.,][]{zaka08} and 2.) the type-2
quasar fraction increases with redshift
\cite[e.g.,][]{hasi08,trei06,lafr05}.

We present here the first systematic search for CO in optically
obscured (type-2) quasars at redshifts 0.09$<$z$<$0.41. The general
properties of the sources in the sample are summarized in
Table~\ref{tab1}. The sources were selected based on their
24$\mu$m-fluxes (MIPS, IRAC) and spectroscopic redshifts from the
type-2 quasar studies conducted by \cite{lacy07b} and \cite{zaka08}.
The upper limit for the redshifts was chosen so that the CO(J=1--0)
line emission still falls within the 3mm band of the IRAM Plateau de
Bure Interferometer (PdBI) receivers. The type-2 quasar sample from
\cite{zaka08} originates from the SDSS catalogue and was followed-up
with Spitzer. The two samples used in \cite{lacy07b} are taken from
the SWIRE- (RA=02h) and the Spitzer XFLS fields (RA=17h).

\begin{figure*}[!t]
\begin{center}
\resizebox{\hsize}{!}{\rotatebox{0}{\includegraphics{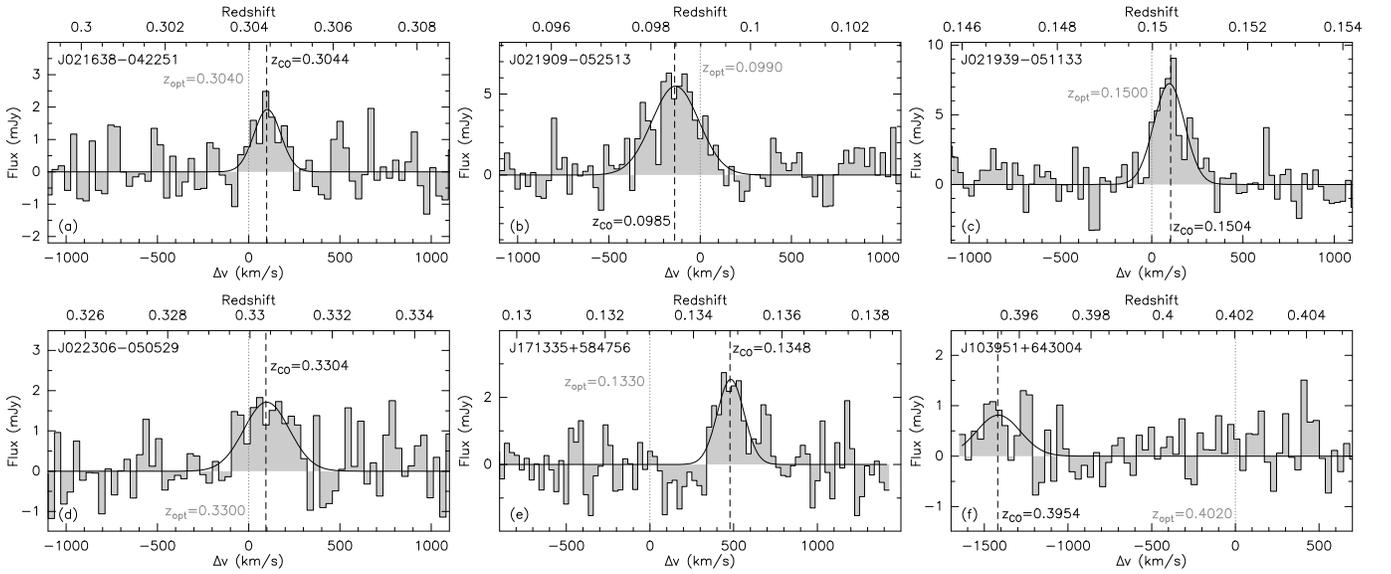}}}
\end{center}
\caption{Spatially integrated CO(J=1--0) line spectrum of all six CO
detections. The zero velocity corresponds to the (spectroscopically
derived) optical redshift ({\it dotted grey lines}). The new redshifts
based on the CO(J=1--0) lines are determined from fitting a Gaussian
profile to the spectra ({\it dashed black lines}).}
\label{fig1}
\end{figure*}

\begin{deluxetable*}{ccccr@{\,}c@{\,}l@{\,}c@{\,}l}
\centering
\tablewidth{0pt}
\tablecaption{Journal of Observations}
\tablehead{
\colhead{Source} & \colhead{Observing Date}
 &  \colhead{Obs. Frequency} & \colhead{Integration time$^a$} & \multicolumn{5}{c}{Synthesized Beam$^b$}  \\
 & (Year:Month) &  (GHz) & (hours) &  \multicolumn{5}{c}{($''$ $\times$$''$@$^\circ$)} }
\startdata 
J021638$-$042250  & \multicolumn{1}{l}{2009:May,Jun,Jul}               &  88.398 &  7.4 &  7.4 & $\times$&5.0 & @ & 153   \\
J021909$-$052512  & \multicolumn{1}{l}{2007:Oct,Nov}                   & 104.887 &  1.9 &  6.9 & $\times$&4.8 & @ & 21    \\  
J021939$-$051133  & \multicolumn{1}{l}{2007:Sep}                       & 100.236 &  4.0 &  6.0 & $\times$&4.7 & @ & 161   \\  
J022306$-$050529  & \multicolumn{1}{l}{2009:May,Jun}                   &  86.670 & 10.3 &  8.3 & $\times$&5.1 & @ & 148   \\
J022508$-$053917  & \multicolumn{1}{l}{2009:Sep,Oct}                   &  89.150 &  5.4 & 11.1 & $\times$&4.6 & @ & 146   \\

J092014+453157    & \multicolumn{1}{l}{2009:May,Jun,Aug; 2010:May,Apr} &  82.206 & 11.3 &  6.2 & $\times$&4.7 & @ & 127   \\
J103951+643004    & \multicolumn{1}{l}{2009:May,Jun; 2010:Apr}         &  82.219 &  7.6 &  6.5 & $\times$&5.4 & @ & 129   \\

J171325+590531    & \multicolumn{1}{l}{2007:Aug}                       & 102.372 &  3.0 &  5.9 & $\times$&3.8 & @ & 74   \\  
J171335+584756    & \multicolumn{1}{l}{2007:Oct; 2009:Sep}             & 101.740 &  8.2 &  5.2 & $\times$&3.3 & @ & 69   \\  
J172123+601214    & \multicolumn{1}{l}{2007:Aug,Sep,Oct}               &  86.997 &  6.5 &  7.2 & $\times$&5.2 & @ & 134  \\   [-0.15cm]
\enddata 

\tablenotetext{$a$}{On-source integration time estimated with respect
to 6 antennas and dual polarisation.}
\tablenotetext{$b$}{For natural weighting.}
\label{tab2}
\end{deluxetable*}

\section{Observations}
The observations of the CO(J=1--0) emission in a sample of ten type-2
quasars (Table~\ref{tab1}) have been conducted at the IRAM PdBI in
2007, 2009 and 2010, using five to six antennas in the most compact
configuration (D) with baselines ranging between 15~m and 100~m. A
journal of the observations is given in Table~\ref{tab2}. We tuned the
3~mm receivers of the PdBI to the redshifted frequency of CO(J=1--0)
(115.271~GHz at rest) for each source, using a bandwidth of 1~GHz
($\simeq$3000-3500~km/s) and a spectral resolution of 2.5~MHz for the
observations dated prior to 2010. In 2010, the new wideband correlator
WideX was installed at the PdBI so that we had a total of 3.6~GHz of
bandwidth at our disposal. Typical system temperatures have been
around $\sim$80-200~K during the observations. We discarded all data
with phase noise exceeding 50~degrees. We calibrated the data using a
set of strong sources, including MWC349 for absolute flux
calibration. The uncertainty of the absolute flux calibration is
estimated to be $\lesssim$10\%.

\begin{figure*}[!t]
\begin{center}
\resizebox{\hsize}{!}{\includegraphics{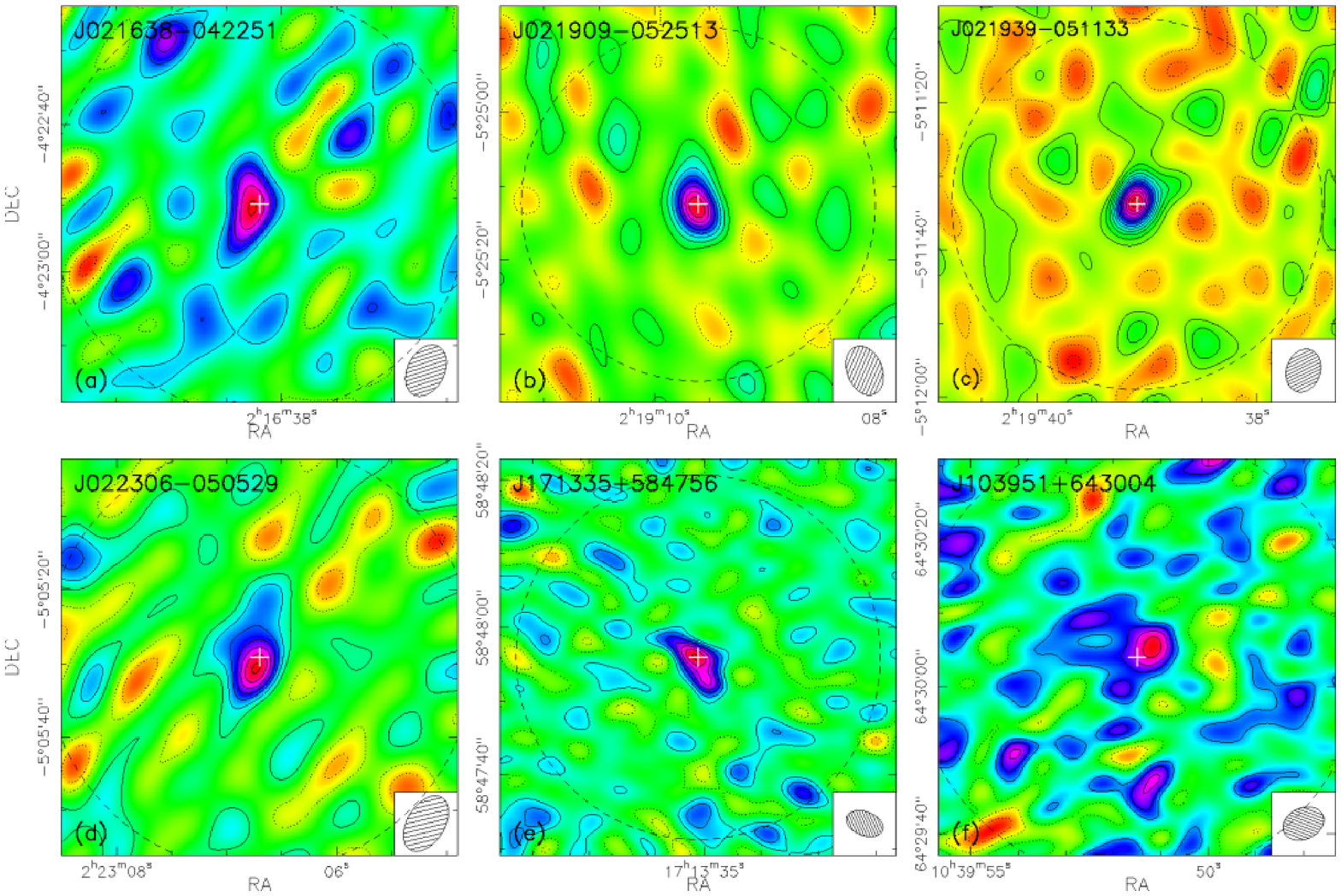}}
\end{center}
\caption{Velocity integrated CO(J=1--0) line emission maps of the six
detected type-2 quasar host galaxies. The large dashed circles
indicate the field of view for each target. The synthesized beams are
shown in the lower right corner of each map. Negative contours
(dotted) correspond to $-$3,$-2$,$-1$$\sigma$, and positive contours
(solid) start at 1$\sigma$ in steps of 1$\sigma$ with the following
noise levels for each map:
(a) 1$\sigma$=0.05~Jy~beam$^{-1}$~km~s$^{-1}$ ($\Delta$v$_{\rm int}$$\simeq$220~km~s$^{-1}$); 
(b) 1$\sigma$=0.58~Jy~beam$^{-1}$~km~s$^{-1}$ ($\Delta$v$_{\rm int}$$\simeq$490~km~s$^{-1}$); 
(c) 1$\sigma$=0.39~Jy~beam$^{-1}$~km~s$^{-1}$ ($\Delta$v$_{\rm int}$$\simeq$330~km~s$^{-1}$); 
(d) 1$\sigma$=0.08~Jy~beam$^{-1}$~km~s$^{-1}$ ($\Delta$v$_{\rm int}$$\simeq$400~km~s$^{-1}$); 
(e) 1$\sigma$=0.08~Jy~beam$^{-1}$~km~s$^{-1}$ ($\Delta$v$_{\rm int}$$\simeq$350~km~s$^{-1}$);
(f) 1$\sigma$=0.07~Jy~beam$^{-1}$~km~s$^{-1}$ ($\Delta$v$_{\rm int}$$\simeq$420~km~s$^{-1}$)}
\label{fig2}
\end{figure*}

\begin{deluxetable*}{cccccccccc}
\centering
\tablewidth{0pt}
\tablecaption{Bolometric (radio, IR, X-ray) properties  of the observed sources}
\tablehead{
\colhead{Source} 
& \colhead{S$_{\rm 160\mu m}$$^a$} & \colhead{S$_{\rm 70\mu m}$$^a$} 
& \colhead{S$_{\rm 24\mu m}$$^a$} & \colhead{S$_{\rm 8\mu m}$$^a$} 
& \colhead{L$_{\rm FIR}$}$^b$ & \colhead{L$_{\rm MIR}$}$^b$ & \colhead{L$_{\rm IR}$}$^b$
& \colhead{S$_{\rm 1.4GHz}$$^{c}$} & S$_{\rm 0.2-12~keV}$$^d$ \\
           & (mJy) & (mJy)      & (mJy) & (mJy) & (10$^{10}$L$_\odot$) & (10$^{10}$L$_\odot$) 
& (10$^{10}$L$_\odot$) & (mJy) & (10$^{-14}$ ergs$^{-1}$ cm$^{-2}$) }
\startdata 
J021638$-$042251    & \nodata &  27      & 14.6    & 2.7     & $>$8        & $\sim$23 & $>$32    & \nodata       & \nodata \\
J021909$-$052513    & 152     & 130      & 25.6    & 3.1     & $\sim$6     & $\sim$3  & $\sim$9  & 1.36$\pm$0.02 & 2.6     \\  
J021939$-$051133    & \nodata &  70      & 32.9    & 8.5     & $\sim$4     & $\sim$11 & $>$15    & 1.24$\pm$0.06 & 5.3     \\  
J022306$-$050529    & \nodata & \nodata  & 15.7    & 1.5     & \nodata     & $\sim$30 & $>$30    & \nodata       & 0.95    \\
J022508$-$053917    & \nodata &  32      &  9.6    & 2.6     & $>$9        & $\sim$14 & $>$23    & \nodata       & \nodata \\
J092014+453157    & \nodata & \nodata  & \nodata & \nodata & \nodata     & \nodata  & $\sim$8  &  \nodata      &   44 \\
J103951+643004    & \nodata & \nodata  & \nodata & \nodata & \nodata     & \nodata  & $\sim$13 &  4.3$\pm$0.5  &   70 \\
J171325+590531    & \nodata & \nodata  &  9.5    & 1.19    & \nodata     & $\sim$2  & $>$2     & 0.16$\pm$0.03 & \nodata \\  
J171335+584756    & 146     &  97      & 23.7    & \nodata & $\sim$9     & $\sim$6  & $\sim$15 & 3.4$\pm$0.2   & \nodata \\  
J172123+601214    & \nodata & \nodata  & 13.3    & 3.71    & $\sim$4$^d$ & $\sim$25 & $\sim$29 & 0.26$\pm$0.04 & \nodata \\   [-0.15cm]
\enddata 

\tablenotetext{$a$}{taken from \cite{lacy04}, \cite{lacy07b},
           \cite{fray06} and/or from the SWIRE catalogue available at {\tt
           http://swire.ipac.caltech.edu/swire/astronomers/data\_access.html} .}

\tablenotetext{$b$}{determined based on L$_{\rm
              FIR}$=$\zeta_2(z)\nu$L$_\nu$(70$\mu$m)+$\zeta_3(z)\nu$L$_\nu$(160$\mu$m),
              L$_{\rm MIR}$=$\zeta_1(z)\nu$L$_\nu$(25$\mu$m), and
              L$_{\rm IR}$=L$_{\rm FIR}$+L$_{\rm MIR}$ from
              \cite{dal02}; with $\zeta_1(z\approx0.1)$$\approx$1.75,
              $\zeta_1(z\approx0.3)$$\approx$2.2,
              $\zeta_2(z\approx0.1)$$\approx$
              $\zeta_2(z\approx0.3)$$\approx$0.82,
              $\zeta_3(z\approx0.1)$$\approx$$\zeta_3(z\approx0.3)$$\approx$1.347. The
              IR luminosities for J092014+453157 and J103951+643004
              were taken from \cite{zaka08}.}

\tablenotetext{$c$}{References: J17-sources: \cite{cond03},
J02-sources: \cite{simp06}, J103951+643004 from \cite{cond98}.}

\tablenotetext{$d$}{from \cite{lacy07b} and \cite{zaka08}.}
\label{tab3}
\end{deluxetable*}

\section{Results}

\subsection{Line Emission}

We detect CO(J=1--0) emission in five of the ten type-2 quasars and
find a tentative detection in a sixth source (see Table~\ref{tab4} and
Fig.~\ref{fig1} \& \ref{fig2}), resulting in a $\sim$60\% detection
rate which is similar to previous CO surveys on quasars
\cite[e.g.,][]{evan01,scov03,bert07}.  The CO(J=1--0) emission in the
remaining four sources remains undetected. 

In the course of our observations, we ``detected'' $\geq$5$\sigma$
emission peaks at off-center positions in the field of view of at
least four of our ten targets. None of these peaks were found to
coincide in position with any of our quasars nor any known galaxy in
the NED data base. Statistical considerations (see Appendix) further
argue for a spurious nature of these $\sim$5$\sigma$ peaks; we
dedicate a section on the Appendix to this problem. Otherwise no
further consideration is given to these noise peaks in the following.

\subsubsection{Detections at the Phase Center}

The line emission of five of our ten targets is clearly detected in
the line spectrum and the integrated maps. The spatial position of the
emission (Fig.~\ref{fig2}) as well as the redshift derived from the
line centers (Fig.~\ref{fig1}) coincide well with the spatial
positions from optical/IR/radio observations
\citep[][Table~\ref{tab3}]{simp06} and their photometric redshifts
(see Table~\ref{tab1} and \ref{tab4}). The CO redshifts were
determined by fitting a Gaussian profile to the line spectrum of each
source.

We further find a tentative detection of the CO(J=1--0) emission in a
sixth source (see Fig.~\ref{fig1} and \ref{fig2}). A 4.5$\sigma$ peak
is found close to the position of the radio emission of this source.
The CO line, however, is significantly shifted bluewards with respect
to what is expected from the photometric redshift. The difference is
much larger than for any of the other five detected sources. Also, the
line profile of the CO(J=1--0) emission is not statistically
significant, so that additional observations would be needed to
confirm or discard this tentative detection. However, since the
4.5$\sigma$ peak is within the radio position of this quasar, we think
that it is reasonable to assume that this detection is indeed real but
we will still treat this source as tentative for the remainder of the
discussions.

All six sources exhibit simple line profiles that appear Gaussian in
nature, single peaked and narrow. To derive the line intensities, line
centers, and line width, we fitted a Gaussian function to all six
lines. The results are given in Table~\ref{tab4}.

\begin{figure*}[!t]
\centering
\rotatebox{0}{\resizebox{\hsize}{!}{\includegraphics{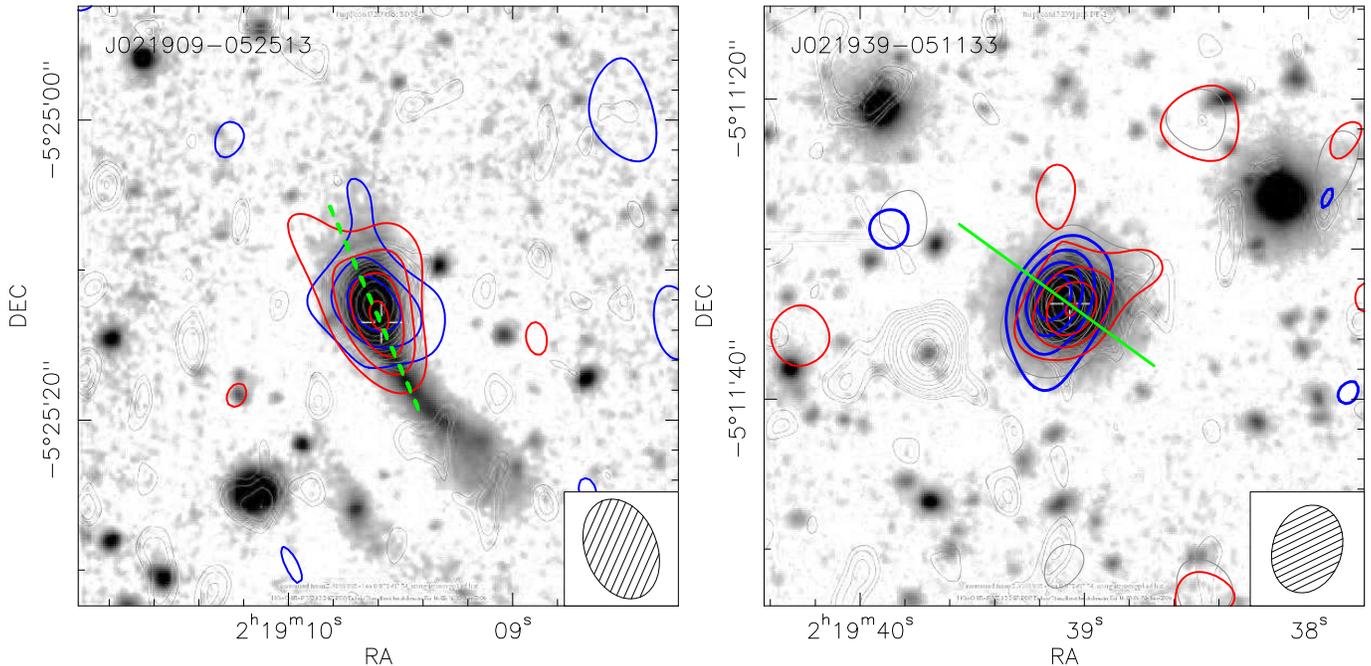}}}
\caption{Blue- and redshifted CO(J=1--0) line emission ({\it
contours}) of J021909$-$052513 ({\it left panel}) and J021939$-$051133
({\it rigt panel}) overlaid to the Subaru/Suprime-Cam i' image ({\it
grey scale}) and the radio emission ({\it faint grey contours}) taken
from \cite{simp06}. The velocity gradient is identicated with a green
line; the green line is dashed for J021909$-$052513 because a velocity
gradient is much more difficult to identify than for
J021939$-$051133. Contours start at 2$\sigma$ and go in steps of
2$\sigma$.}
\label{fig3}
\end{figure*}

Four out of these six galaxies appear to be unresolved at their
detection level and no significant velocity gradient can be
identified. Only the CO emission regions of J021909$-$052513 and
J021939$-$051133 appear to be slightly extended and show a velocity
gradient along a North-East to South-West direction (Fig.~\ref{fig3})
although the one for J021909$-$052513 is less pronounced. While the
optical i'-image \cite[taken from][]{simp06} shows a tail-like
extension for J021909$-$052513 indicating a possible interaction with
another galaxy, the optical image for J021939$-$051133 shows a compact
galaxy with no signs of disruption.  The CO emission in
J021909$-$052513 suggests the presence of molecular gas along the
tidal tail of this merger system. Fitting a Gaussian to the uv-data
results in a size estimate of 2$\pm$1$''$ ($\sim$5~kpc), i.e., very
centralised molecular gas. However, at the angular resolution and
sensitivity of our observations, we cannot exclude that part of the
more extended diffuse gas might have been resolved out.

\begin{deluxetable*}{c@{}c@{}c@{}c@{}c@{}c@{}c@{}c@{}c@{}c}
\centering
\tablewidth{\hsize}
\tabletypesize{\scriptsize}
\tablecaption{Observational Results}
\tablehead{
\colhead{Source} &  \colhead{RMS$^a$}    & \colhead{f$_{\rm cont}$$^b$} & \colhead{f$_{\rm CO}$$^{c,d}$}
 & \colhead{$\Delta v_{\rm FWHM}$$^{e}$}  & \colhead{$v_0$$^{f}$}&  $z_{\rm CO}$$^g$  
& \colhead{S$_{\rm CO}$$^d$}  & \colhead{L'$_{\rm CO}$$^{d}$}    & \colhead{M$_{\rm gas}$$^{d,h}$}  \\
                 &  (mJy)  & (mJy) &  (mJy) & (km~s$^{-1}$) & (km~s$^{-1}$) &  & (Jy~km~s$^{-1}$)  
& (K~km~s$^{-1}$~pc$^2$) &  (10$^{9}$M$_\odot$)    }
\startdata 
\multicolumn{10}{c}{DETECTIONS} \\
J021638$-$042251    & 0.8 & $\leq$0.3 & 2.0$\pm$0.4 & 170$\pm$40  & $+$102$\pm$17 & 0.3044$\pm$0.0002 & 0.5$\pm$0.1    & (23$\pm$5)$\times$10$^8$  & 11$\pm$2     \\
J021909$-$052513    & 1.4 & $\leq$0.6 & 5.6$\pm$0.6 & 300$\pm$40  & $-$135$\pm$16 & 0.0985$\pm$0.0006 & 1.5$\pm$0.3    & ( 7$\pm$1)$\times$10$^8$  &  3.2$\pm$0.6 \\
J021939$-$051133    & 1.6 & $\leq$0.5 & 7.3$\pm$0.7 & 200$\pm$30  & $+$100$\pm$10 & 0.1504$\pm$0.0002 & 1.6$\pm$0.2    & (17$\pm$2)$\times$10$^8$  &  8$\pm$1     \\
J022306$-$050529    & 0.7 & $\leq$0.2 & 1.7$\pm$0.3 & 290$\pm$60  &  $+$98$\pm$27 & 0.3304$\pm$0.0004 & 0.6$\pm$0.1    & (33$\pm$5)$\times$10$^8$  & 16$\pm$3     \\
J171335+584756      & 0.8 & $\leq$0.3 & 2.5$\pm$0.4 & 190$\pm$30  & $+$480$\pm$10 & 0.1348$\pm$0.0003 & 0.45$\pm$0.08  & ( 5$\pm$1)$\times$10$^8$  &  2.4$\pm$0.4 \\[0.2cm]

\hline\\[-0.15cm]
\multicolumn{10}{c}{TENTATIVE DETECTIONS} \\[0.1cm]
J103951+643004      & 0.6 & $\leq$0.2 & 0.81$\pm$0.26 & 310$\pm$120  & $-$1415$\pm$50  & 0.3954$\pm$0.0006 & 0.25$\pm$0.06  & (21$\pm$5)$\times$10$^8$  & 10$\pm$2     \\[0.2cm]
\hline\\[-0.15cm]
\multicolumn{10}{c}{NON-DETECTIONS} \\[0.1cm]
J022508$-$053917    & 0.3 & $\leq$0.3 & $\leq$0.9   & \nodata     & \nodata       &  \nodata          & $\leq$0.3$^i$  & $\lesssim$13$\times$10$^8$ & $\leq$6 \\
J092014+453157      & 0.2 & $\leq$0.2 & $\leq$0.6   & \nodata     & \nodata       &  \nodata          & $\leq$0.2$^i$  & $\lesssim$17$\times$10$^8$ &$\leq$8  \\ 
J171325+590531      & 0.5 & $\leq$0.5 & $\leq$1.5   & \nodata     & \nodata       &  \nodata          & $\leq$0.4$^i$  & $\lesssim$3$\times$10$^8$  & $\leq$2 \\
J172123+601214      & 0.2 & $\leq$0.2 & $\leq$0.6   & \nodata     & \nodata       &  \nodata          & $\leq$0.2$^i$  & $\lesssim$11$\times$10$^8$ & $\leq$5 \\[-0.15cm]
\enddata 

\tablenotetext{$a$}{RMS noise determined from channels with 10~MHz
spectral resolution ($\simeq$28~km~s$^{-1}$) for the CO detections and
with 105~MHz spectral resolution ($\simeq$300~km~s$^{-1}$) for the CO
non-dectections.}

\tablenotetext{$b$}{3$\sigma$-upper limits for the continuum flux
averaged over the line-free channels for each source.}

\tablenotetext{$c$}{Peak line fluxes for the detections and
3$\sigma$-upper limits for the non-detections.}

\tablenotetext{$d$}{Errors are of pure statistical nature and do not
account for uncertainties from the absolute flux calibration. The
latter is estimated to be at a conservative level of $\sim$10\%. The
integrated flux densities have been determined assuming that the
continuum emission in these sources is negligible at the given noise
level (see also text).}

\tablenotetext{$e$}{$\Delta v_{\rm FWHM}$= Full Width at Half Maximum
of the line. Determined from CO line profile by fitting a Gaussian to
the data.}

\tablenotetext{$f$}{Systemic velocity; zero velocity corresponding to
the optical redshifts of the sources.}

\tablenotetext{$g$}{Redshift determined from the central velocity of
the CO(J=1--0) line.}

\tablenotetext{$h$}{The standard galactic M$_{\rm gas}$-to-L'$_{\rm CO}$
 conversion factor of X$_{\rm
 CO}$=4.8M$_\odot$~(K~km~s$^{-1}$~pc$^2$)$^{-1}$ has been assumed.}

\tablenotetext{$i$}{Assuming a $\Delta v_{\rm FWHM}$ of
$\sim$260~km~s$^{-1}$ corresponding to the average of the FWHM of the
detected sources.}

\label{tab4}
\end{deluxetable*}

\subsubsection{Non-detections}
Although we cannot entirely rule out this possibility, we doubt that
the four non-detections are due to uncertainties in the optical
spectroscopic redshifts. \cite{hain04} present a comparison between
optical and CO redshifts in high-redshift sources . The authors find,
excluding a few outliers, a difference of $\Delta$z$_{\rm
max}$=$\mid$z$_{\rm opt}$-z$_{\rm CO}\mid\lesssim$0.008, which
corresponds to a velocity shift of $\sim$$\pm$(1500-2000)~km~s$^{-1}$
at 3.5\,mm in the redshift range z=0.2-0.5. A similar deviation of
$\sim$0.008 is seen when adding the CO detections at high-z since 2005
\citep[e.g.,][]{grev05,copp08}. Interestingly, the outliers which show
0.008$<$$\Delta$z$\lesssim$0.05 all lie at redshift
z$\geq$2.5. Therefore, the large bandwidth of the observations of
3000~km~s$^{-1}$ should be sufficiently large to take out
uncertainties in the optical redshift for the sources presented in
this paper. However, it has to be pointed out that for the
determination of the difference $\Delta$z only sources with {\it
detected} CO emission were considered. The actual deviation $\Delta$z
could be hence larger than assumed here.  Only larger bandwidth
($\geq$4~GHz) and higher sensitivity (going down to M$_{\rm
gas}$$\lesssim$10$^8$M$_\odot$ where M$_{\rm gas}$=M(H$_2$+He) is the
total gas mass) observations will allow us to reliably eliminate all
uncertainties from optical redshifts for the detection of CO emission
in higher-z sources.

\subsection{Continuum Emission}
Continuum emission at 3~mm has not been found in any of the ten
sources. The 3$\sigma$ upper limits derived for the 3~mm continuum
fluxes (Table~\ref{tab4}) are similar or slightly lower than the
1.4~GHz flux densities (Table~\ref{tab3}) excluding a flat radio
spectrum in these sources and pointing rather towards a steep radio
spectrum.  A steep radio spectrum is generally associated with
non-thermal emission which is typically found in this galaxy
population. However, extrapolating the 1.4~GHz flux densities
(Table~\ref{tab3}) with a spectral index of $-$1 (i.e., f$_{\rm
\nu}$$\propto$$\nu^{-1}$) infers flux densities at 3~mm that are
smaller by an order of magnitude than our 3~sigma upper limits. We
also stacked the continuum maps from all ten targets with no
significant detection in the stacked image, i.e., the emission peak in
the center remains at $\lesssim$3$\sigma$=0.09~mJy. Thus, no definite
conclusion can be drawn on the nature of the 3~mm continuum emission
in our sources.

\section{Discussion and Conclusions}

\subsection{Molecular Gas Masses}
Using the standard Galactic conversion factor for the M$_{\rm
gas}$-to-L'$_{\rm CO}$ ratio of X$_{\rm
CO}$=4.8~M$_\odot$~(K~km~s$^{-1}$~pc$^2$)$^{-1}$
\cite[e.g.,][]{solo91}, we find a range of the total gas mass of
M$_{\rm gas}$$\simeq$(2-16)$\times$10$^{9}$M$_\odot$ for all five
detected sources (plus the one tentatively detected) and 3$\sigma$
upper limits for the four non-detections of M$_{\rm
gas}$$\lesssim$(2-8)$\times$10$^9$M$_\odot$
(Table~\ref{tab4}). However, the standard Galactic conversion factor
X$_{\rm CO}$ is known to over-estimate the total gas mass in the
ULIRG/submillimeter population by more than a factor of 5
\citep[e.g.,][]{down98} and seems to be also highly dependent on the
physical properties \citep[e.g.,
metallicity;][]{bell07a,bell07b}. \cite{bell07b} find indications for
a lower X$_{\rm CO}$ in AGN dominated galaxies in the local universe
as well (e.g., M51). If this were also true for higher-z quasars, the
gas masses for the sources in this paper would reduce to values below
3$\times$10$^9$M$_\odot$.

\subsection{Dynamical Masses}
Assuming a radius of the emission equal or lower than half the
synthesized beam (i.e., $\sim$3-5$''$$\simeq$5-15~kpc) and taking the
full width at half maximum (FWHM) of the line profiles determined for
the six detected sources, the dynamical masses amount to
$\sim$(10$^{10}$-10$^{12}$)M$_\odot$$\times$sin$^{-2}$$i$ with $i$
being the disk inclination angle. These dynamical masses easily
account for the obtained total gas mass within the assumed radius of
5-15~kpc ($\simeq$ 10\% of the stellar amount in the entire galaxy,
which is a few times 10$^{12}$M$_\odot$) and leave enough room for
additional material, such as the super-massive central black hole (up
to $\sim$10$^{10}$M$_\odot$ in the most extreme cases) and the stellar
content (up to a few times 10$^{11}$M$_\odot$).

\subsection{CO versus IR Luminosities}
\begin{figure*}[!t]
\centering
\rotatebox{0}{\resizebox{\hsize}{!}{\includegraphics{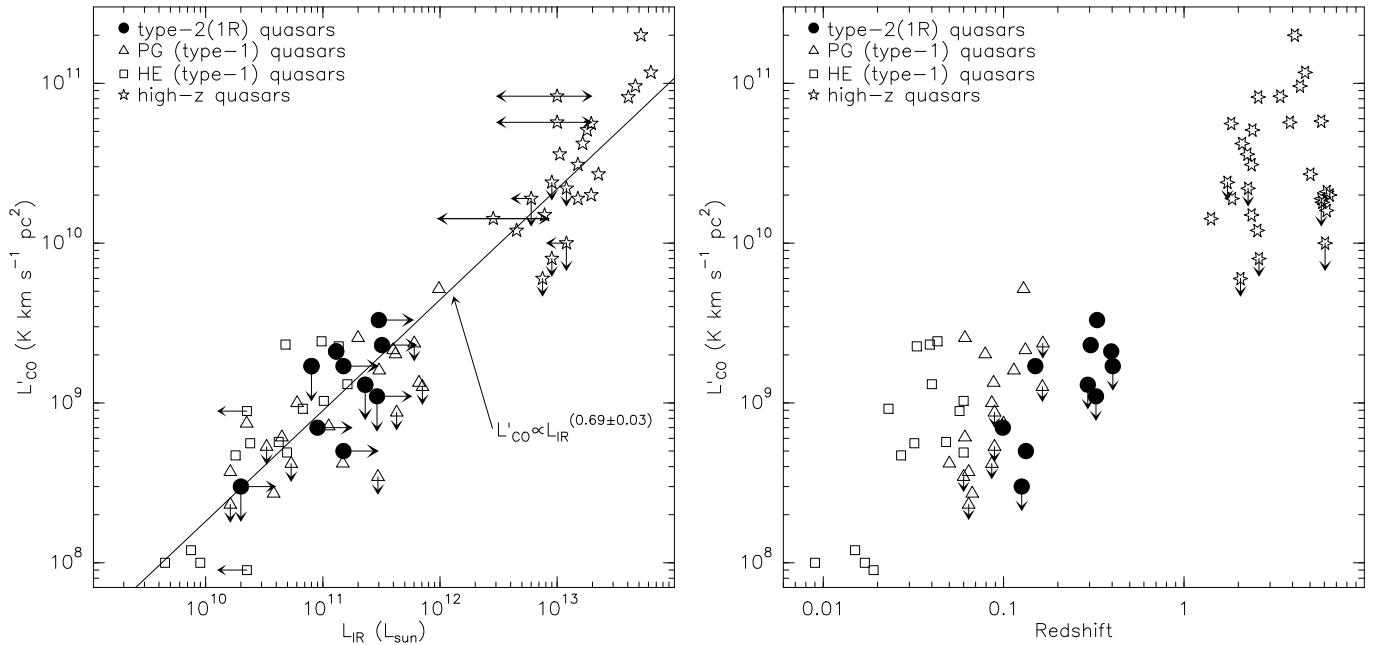}}}
\caption{CO versus IR luminosity plot ({\it left}) and CO luminosity
versus redshift plot ({\it right}) of quasars with CO measurements
taken from this paper (filled black circles) and the literature (open
symbols). The solid black curve ({\it left}) is a linear fit to the
data. Given their high dust obscuration, the IR-luminosities of the
type-2(1R) quasars can only be regarded as lower limits. Upper and
lower limits are marked with arrows. For some of the high-redshift
sources (z$\geq$1), only FIR luminosities were available, so we assume
L$_{\rm IR}$=1.5$\times$L$_{\rm FIR}$ based on a L$_{\rm
FIR}$-to-L$_{\rm IR}$ comparison done by \cite{pott06}; please note
that some quasars can show a ratio larger by a factor of 2-4 because
of the contribution of the AGN to the MIR luminosity. However,
increasing the ratio to 4 does not significantly change the slope of
the correlation and is taken into account in the error of the fit. The
references for the literature data are as follows: PG quasars --
\cite{evan01,scov03}; HE quasars -- \cite{bert07}; high-z quasars --
\cite{wanga11,wangb11,poll11,copp08,ara08,maio07,krip05b,wal04,car02,cox02}.}
\label{fig4}
\end{figure*}

Comparing the CO-luminosities for the detected sources and the upper
limits for the undetected ones with those of type-1 quasars taken from
the literature \citep[e.g.,][]{evan01,scov03,bert07}, we find that
they fall within the same area in the L'$_{\rm CO}$-L$_{\rm IR}$
diagram (Fig.~\ref{fig4}). However, it has to be noted that most of
the IR luminosites assumed for the type-2(1R) quasars must probably be
regarded as lower limits (see Table~\ref{tab3}). Several of our type-2
quasars have no measured/detected FIR fluxes. Also, obscuration
through dust might significantly reduce the emission at MIR
wavelengths. However, based on comparison with other dust enshrouded
quasars, the IR-luminosities can be larger by up to an order of
magnitude than the MIR-luminosities reducing hence the significance of
the MIR to the total IR-luminosities.

Taking all published CO detections in quasars into account and
excluding the upper and lower limits for some of the CO and IR
luminosities, we find a linear correlation between the CO and IR
luminosities of 0.70$\pm$0.03~dex (Fig.~\ref{fig4}). When considering
the upper limits for the CO luminosities and, separately, the lower
limits for the IR luminosities, we obtain a very similar slope of
0.69$\pm$0.03~dex based on the analytical method described in
\cite{feig85}. This is in good agreement with results found for ULIRGs
and nearby active galaxies \cite[e.g.,][]{gao04,nara08a}. However,
some of the upper limits of the CO observations fall significantly
below the correlation but it is unclear whether this is due to a
breakdown in the correlation, to uncertainties in the optical redshift
and/or the significant contribution of the AGN to the
(M)IR-luminosities that might even dominate over that from
star-formation \cite[e.g.,][]{lacy07b,bran06}. Besides obscuration
effects, the latter represents an important factor in determining
``accurate'' (i.e., purely star-formation related) IR-luminosities in
particular for type-2 quasars. However, the significance of the AGN
contribution to the energy output remains also strongly debated for
ULIRGs \cite[e.g.,][]{fran03,iman07} so that the importance of the AGN
influence on the correlation between CO and IR-luminosities is hard to
quantify.  On the other hand, the similarity of this correlation
between ULIRGs and quasars may indicate that either the AGN
contribution influences the correlation in the same way for these
sources (such as adding an offset in a systematic way) or is even a
fundamental part of the correlation. A detailed entangling of the
different components is certainly necessary to further our
understanding of this very fundamental correlation.

Alternatively, the slope of the L'$_{\rm CO}$-L$_{\rm IR}$ correlation
could also vary for different galaxies \citep[as discussed
by][]{gao04,bert07}. \cite{bert07} suggests a scenario in which
geometrical effects naturally lead to a change in slope: galaxies with
highly centralised and confined molecular gas (such as ULIRGs) have a
larger surface filling factor and thus a steeper slope whilst 'normal'
weakly interacting/active galaxies show more extended molecular gas
resulting in a milder slope. The similarity of the slope between
ULIRGs and quasars might hence be an indication that merger events are
also important for the quasar population. However, more observations
and studies of the distribution of CO in quasars, and active galaxies
in general, have to be undertaken to validate this interpretation.

\subsection{CO Luminosities versus redshift}
Plotting the CO luminosities of the quasars detected in CO against
their redshifts (right panel of Fig.~\ref{fig4}), suggests that,
despite similar IR luminosities, type-2 quasars have lower CO
luminosities, and hence a lower amount of molecular gas at comparable
stages of evolution than type-1 quasars, i.e., similar redshifts; this
impression is mainly given when looking at redshifts between 0.08 and
0.2 (above redshifts of 0.2 and below redshifts of 1.0 no CO detection
for type-1 quasars has yet been reported). When applying the
stastistical analysis from \cite{feig85} that includes the upper
limits (see also previous Section) we do find marginally different
mean values for the CO luminosities in type-1 quasars of $<$L'$_{\rm
CO}$$>$=(13$\pm$4)$\times$10$^8$ (K~km/s~pc$^2$) and those in type-2
quasars of $<$L'$_{\rm CO}$$>$=(8$\pm$3)$\times$10$^8$ (K~km/s~pc$^2$)
for redshifts between 0.08 and 0.2; for the entire type-1 and type-2
quasar samples we find almost the same mean values of $<$L'$_{\rm
CO}$$>$=(12$\pm$3)$\times$10$^8$~(K~km/s~pc$^2$). However, not only do
the mean values agree within the errors, a logrank analysis
\cite[][]{feig85} also shows that both samples are similar at a 50\%
significance level. Therefore we can neither rule out that both
samples exhibit similar CO luminosities nor can we support it with the
current samples.

Even if the CO luminosities were proven to be lower in type-2 than
type-1 sources this would be at odds with both unification
theories. The nature of the viewing-angle unification theory suggests
that no statistically relevant differences should exist between CO
luminosities of type-1 and type-2 quasars. On the other hand, if the
merger-driven unification were to hold true, higher CO luminosities
(i.e., higher molecular gas masses) would be expected for type-2
quasars.

Thus, different explanations must be considered. One would be the low
number statistics. An alternative reason could be the different
selection criteria used for the CO surveys of the type-1 and type-2
quasars. It seems that the type-1 quasars were chosen based on their
bright IR luminosities while this wasn't a criterium for the type-2
sources. The type-1 quasar sample might hence be biased toward gas
rich sources. Also, we cannot exclude resolution effects and hence an
underestimation of the CO luminosities in our interferometric
study. However, the interferometric observations carried out by
\cite{evan01} and \cite{scov03} probably suffered from similar
problems given the similar redshift ranges of the targets and hence
similar spatial scales. Only larger samples will help to settle the
question whether the CO luminosities are different for the two quasar
populations.

\begin{figure}[!t]
\centering
\rotatebox{0}{\resizebox{8cm}{!}{\includegraphics{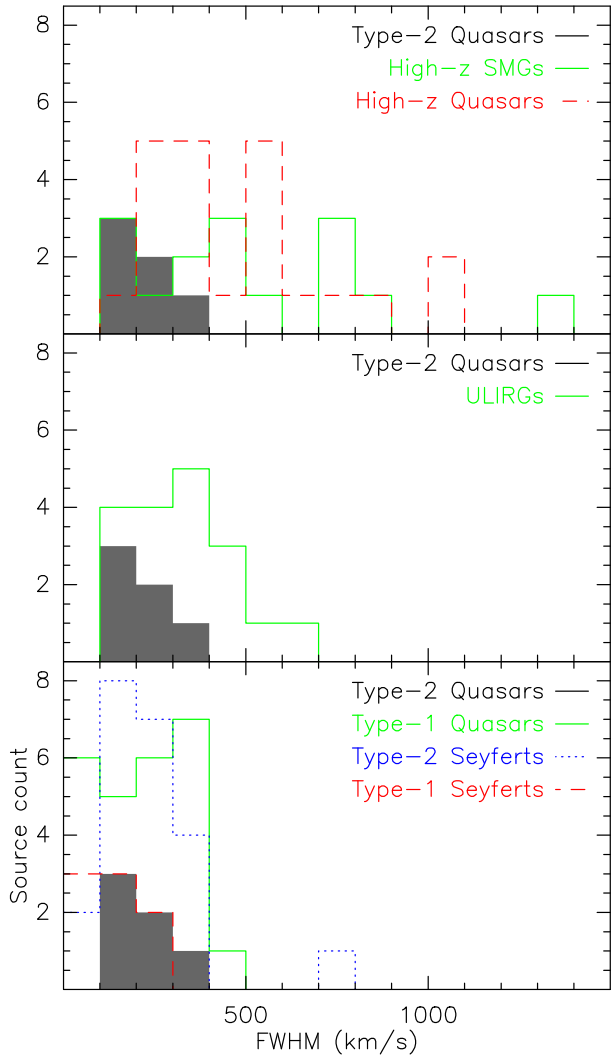}}}
\caption{A comparison of the measured CO linewidth of the six detected
type-2 quasars to other active galaxies in the local and high-redshift
universe; the histogramms were derived such as that ${\rm FWHM}_{\rm
min}<{\rm FWHM}\leq{\rm FWHM}_{\rm max}$. The FWHM for the ULIRGs and
high-redshift galaxies are taken from references given in
Fig.~\ref{fig4} and \cite{iono09}, for the nearby Seyfert galaxies and
type-1 quasars from
\cite{barv89,allo92,vila98,saka99,evan01,scov03,krip05a,bert07}.  {\it
Note of caution: not corrected for inclination; the CO for the high-z
sources is mostly observed at J$_{\rm upper}>$1 so that their FWHM
might be even higher.}}
\label{fig5}
\end{figure}

\subsection{Linewidth of CO Emission}
In case of a lack of sufficient angular resolution, the linewidth of
molecular lines can be a first-order merger-activity indicator for
galaxies, i.e., the larger the linewidth the more likely a galaxy
might be interacting or merging with another galaxy \cite[e.g.,
Arp~220; see also][]{grev05}; however, the contrary is not necessarily
true. Comparing hence the linewidths of our type-2 quasars with that
of ULIRGs and high-z galaxies (Fig.~\ref{fig5}), for which a high
percentage is assumed to be in a merger stage, shows indeed that the
distribution of the linewidth peaks at higher values for the
ULIRG/high-redshift galaxy population than for the nearby type-2
quasars. Also, the ULIRG/high-redshift galaxy population exhibits some
very large linewidths with values exceeding 500~km/s as opposed to the
type-2 quasars for which all measured linewidths stay well below
400~km/s. No clear difference though can be identified between the
type-2 and type-1 quasars; the same is also true for the more local
type-1 and type-2 Seyfert galaxies. Both quasars and Seyferts reveal a
similar distribution of their linewidths. A significant difference in
the distribution of the linewidth between the type-1 and type-2
quasars could favor a merger-driven unification theory; the broadest
linewidths should occur for pre-coalescent mergers for which the gas
disks of the merging galaxies have not yet met
\citep[e.g.,][]{iono09}.

Even in the viewing-angle unification, one might expect a difference
between the linewidths for type-1 and type-2 quasars under certain
assumptions. Let us assume that the molecular gas is mostly located in
a disk in these quasars and that this (large-scale) disk exhibits a
similar inclination to the putative (small-scale) torus. If that were
true, the CO emission would be mostly seen edge-on in type-2 and
face-on in type-1 quasars. In turn this means that the CO linewidth
would generally be larger for type-2 than for type-1 quasars. However,
overall the linewidths are not expected to exceed values of
500~km~s$^{-1}$ as found in mergers. Therefore, a distinction to the
merger unification should still be possible. On the other hand, the
assumptions made above may not be valid. If a merging event is indeed
taking place in these systems, the molecular gas might be located in a
severely disrupted disk for which an inclination would be difficult to
derive. Also, even if located in a well defined (large scale) disk,
its inclination does not have to be the same as that of a small scale
torus \cite[see for instance the discussion for M51 and NGC~1068
in:][]{mats07,krip11}.

However, one has to keep in mind that there are a few caveats to be
considered for this comparison.  First, the size of the sample suffers
from small number statistics. Second, no correction has been done for
disk inclination effects, but, since this is true for all galaxy
populations used here, the effects are assumed to be on average the
same for all and to cancel each other out. Third, the CO lines for the
high-redshift galaxy population have been detected in higher
rotational transitions which can underestimate the actual linewidth as
they tend to trace the warmer and more compact molecular gas. Fourth,
also the possible presence of gas outflows is to be considered a
linewidth broadening factor
\cite[e.g.,][]{nara06,nara08b,nara08c,greve08,feru10}.

\section{Summary}

We detected the CO(J=1--0) line emission in five out of a sample of
ten type-2 quasars. We further report on the tentative detection of
the CO(J=1--0) line emission in a sixth source. The estimated
molecular gas masses are rather moderate with values not exceeding
2$\times$10$^{10}$M$_\odot$. All detections nicely follow the L'$_{\rm
CO}$-to-L$_{\rm IR}$ correlation for quasars. The CO luminosities
derived for the six detections and the upper limits for the remaining
four type-2 quasars seem to lie slightly below that of previously
detected type-1 quasars for similar redshifts. However, given the low
number statistics of the current samples this difference is not found
to be at a statistically significant level and might simply be due to
a selection bias. The rather simple CO(J=1--0) line profiles and
moderate CO luminosities do not seem to support a recent or current
merger activity in these sources; however, at least in the case of one
source, J021909$-$052513, the presence of a tidal tail in the optical
images is strongly suggestive of a recent or on-going merger
event. The data discussed in this paper do not allow to favor either
the merger-driven or the viewing angle unification models. Higher
number statistics are mandatory to support one or the other scenario.


\acknowledgments

\appendix

\begin{figure*}[!t]
\centering
\resizebox{\hsize}{!}{\rotatebox{0}{\includegraphics{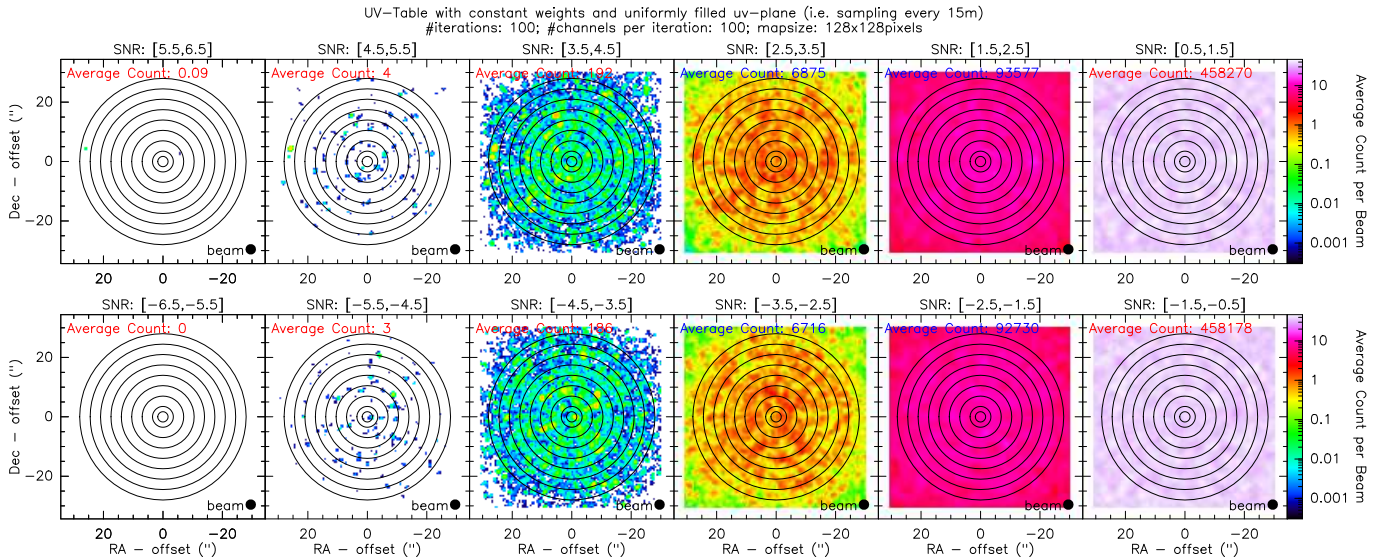}}}
\caption{Statistical analysis of the Gaussian noise peak distribution
for a uniformly filled uv-coverage with one visibility of constant
weight every 15m. 100 independent simulations for a 100-channel data
cube have been run (for an observing frequency of $\sim$100~GHz in D
configuration of the PdBI). The maps show the average number count per
pixel for each peak intensity normalised by the beamsize within a
certain $\sigma$-range. The sum of the total number of peaks per map
divided by the number of repetitions and normalised by the beamsize is
given in the upper left corner of each map. The top panels show the
positive peaks, the bottom panels the negative ones. The circles
indicate radii starting at 0.5$\times$$\theta_s$, $\theta_s$ and then
increasing in steps of $\theta_s$ ($\theta_s$=3.5$''$). The
synthesised beam $\theta_s$ is shown in the lower right corner.}
\label{fig6}
\end{figure*}

\section{Detections of $\sim$5$\sigma$ peaks at off-center positions}

The CO detections and non-detections for the ten sources were
determined by running an automated procedure on the calibrated data
cubes in both the uv- and the (dirty) image plane. This procedure
generates integrated maps for all consecutive channel combinations,
i.e., for a 100-channel data cube the code integrates channel number 1
to 2, 1 to 3, ..., 1 to 100, 2 to 3, 2 to 4, 2 to 100, until it
reaches 99 to 100. It then searches for the emission maximum in each
of the integrated maps and derives the absolute maximum of those
individual maxima. This provides a robust and unbiased approach in the
search for line emission in large spectral data cubes. With this
method several $\geq$5$\sigma$ line emission peaks were observed in
addition to the detections and non-detections at the center of the
arrays field of view.

The uncertainties of the optical/IR positions of the type-2 quasars
are smaller than 1$''$. This is much smaller than the difference
between the actually targeted type-2 quasars and the detected emission
peaks which were found to be off by more than $\sim$5$''$ from the
quasar position ($\geq$2-10$\theta_s$). A positional search close to
the peaks using NED did not show any other galaxies that could be
associated with these peaks raising doubts about the reliability of
these detections.

Although a $\geq$5$\sigma$ peak in the velocity integrated intensity
map is generally assumed to mark a reliable detection limit for line
emission with the IRAM PdBI, even for off center positions, the
significant increase in spectral bandwidth might necessitate a
refinement of this detection limit based on statistical
considerations. The probability of finding several consecutive
2-3$\sigma$ noise peaks in adjacent spectral channels at any position
within (1.5$\times$) the primary beam increases with the spectral
bandwidth and ultimately leads to spurious detections. In order to
assess the probability of spurious detections, we conducted a
statistical analysis of the data based on an ideal noise distribution
as a starting point. Ideal in this context stands for a uniformly
filled uv-plane (i.e., one visibility with a constant weight every 15m
in either direction up to the maximum D-configuration baseline of 100m
used in the actual observations) and a Gaussian noise distribution. We
used 100 channels per simulated data cube with 128$\times$128 pixels
per channel map. We ran the aforementioned ``peak-finder'' procedure
and repeated this 100 times, every time with a re-computed noise
distribution. The results are shown in Fig.~\ref{fig6}. A peak-count
map is shown for different levels of detection in units of $\sigma$,
i.e., as multiples of the rms noise (starting from 0.5$\sigma$ up to
6.5$\sigma$ peaks).  The upper panel shows the average counts of
positive peaks per pixel normalised to the beamsize while the lower
panel that of the negative ones. The absolute average count that is
derived from the total counts per map divided by the number of
iterations (100) and normalised by the beamsize ($\sim$30~pixel$^2$
per beam) is plotted in each SNR range map. The circles indicate
different radii, starting at r=$\theta_2$/2,$\theta_2$ and increasing
then in steps of $\theta_s$; in our example the synthesized beam has a
size of $\theta_s$=3.5$''$ (for an observing frequency of
$\sim$100~GHz in D configuration).

Even in the ideal case of uncorrelated noise, constant visibility
weights and even sampling of the uv-plane, we obtain, on average, at
least one $\sim$5$\sigma$ peak anywhere in (1.5$\times$) the primary
beam. Even the probability to find a peak with $>$5.5$\sigma$ is
non-zero and lies at the $\sim$9\% level. In other words, for a CO
line survey of 10 sources this hence means that one spurious
$\geq$5.5$\sigma$ peak will appear within the field of view of at
least one of the ten sources. Furthermore, close to the center of the
field of view, i.e., within a radius equal to the size of $\theta_s$,
the probability to find a $\sim$5$\sigma$ peak is estimated to be at
the $\sim$5\% significance level. Decreasing this circle however
further below a radius of $\sim$1.75$''$ ($\equiv$5 times the
positional uncertainty $\sigma_{\rm p}$ for a synthesized beam of
$\theta_{\rm s}$=3.5$''$ and a signal to noise ratio of SNR=5$\sigma$
assuming $\sigma_{\rm p}$=$\theta_{\rm s}$/(2SNR)), the probablility
drops to almost zero to detect a noise peak with $\geq$4.5$\sigma$,
giving statistical support for the tentative detection of the
CO(J=1--0) emission in J103951+643004.

Obviously, an uneven sampling of the uv-plane, systematic calibration
uncertainties, and correlated noise between adjacent channel maps and
pixels is likely to increase the number of $>$4.5$\sigma$ peaks in the
field of view. Spurious detections will certainly challenge future
detection surveys as bandwidths will steadily increase in the coming
years, especially in the context of a blind survey for line emission
from galaxies at high redshift. Based on these statistics, we repeated
the observations for those sources for which we found $\geq$5$\sigma$
off-center peaks within the field of view. None of these peaks could
be confirmed to be real.



\end{document}